\documentclass{ar-1col-S2O}
\usepackage[numbers]{natbib}
\jname{Xxxx. Xxx. Xxx. Xxx.}
\jvol{AA}
\jyear{YYYY}
\doi{10.1146/annurev-nucl-121423-101050}

\usepackage{xspace}
\usepackage{amsmath,amsfonts,amssymb} 
\usepackage[colorlinks]{hyperref}
\usepackage{heppennames}

\usepackage{lmodern}

\newcommand{\mult}{\ensuremath{\langle \text{d}N_{\text{ch}}/\text{d}\eta \rangle}\xspace}
\newcommand{\pT}{\ensuremath{p_{\text{T}}}\xspace}
\newcommand{\kT}{\ensuremath{k_{\text{T}}}\xspace}
\newcommand{\meanpT}{\ensuremath{\langle p_{\text{T}} \rangle}\xspace}
\newcommand{\pTzero}{\ensuremath{p_{\text{T}0}}\xspace}
\newcommand{\xT}{\ensuremath{x_{\text{T}}}\xspace}
\newcommand{\gevc}[1]{\ensuremath{#1 \text{\,GeV/$c$}}\xspace}
\newcommand{\KZ}{\ensuremath{K^{0}_{s}}\xspace}
\newcommand{\LA}{\ensuremath{\Lambda}\xspace}
\newcommand{\XI}{\ensuremath{\Xi}\xspace}
\newcommand{\XIM}{\ensuremath{\Xi^{-}}\xspace}
\newcommand{\OM}{\ensuremath{\Omega}\xspace}
\newcommand{\PHI}{\ensuremath{\phi}\xspace}
\newcommand{\LQCD}{\ensuremath{\Lambda_{QCD}}\xspace}

\newcommand{\ee}{\ensuremath{e^+e^-}\xspace}
\newcommand{\qqbar}{\ensuremath{q\overline{q}}\xspace}

\newcommand{\pythia}{\textsc{Pythia}\xspace}
\newcommand{\epos}{\textsc{Epos}\xspace}
\newcommand{\eposlhc}{\textsc{Epos Lhc}\xspace}
\newcommand{\herwig}{\textsc{Herwig}\xspace}
\newcommand{\rivet}{\textsc{Rivet}\xspace}
\newcommand{\professor}{\textsc{Professor}\xspace}
\newcommand{\mcplots}{\textsc{MCPlots}\xspace}

\begin{document}

\markboth{P.\@ Christiansen and P.\@ Van Mechelen}{Soft QCD at LHC}

\title{Soft QCD Physics at the LHC: highlights and opportunities}

\author{P.~Christiansen,$^1$ and P.~Van~Mechelen$^2$
\affil{$^1$Department of Physics, Lund University, Lund, Sweden, SE-22362; email: peter.christiansen@fysik.lu.se}
\affil{$^2$Department of Physics, Antwerp University, Antwerp, Belgium, B-2020;
  email: pierre.vanmechelen@uantwerpen.be}}

\begin{abstract}
The Large Hadron Collider (LHC) at CERN in Switzerland became operational in 2009 and has since then produced a plethora of results for proton-proton (pp) collisions. This short review covers results that relates to soft QCD focusing on non-diffractive physics at mid-rapidity. Most of the presented results comes from transverse momentum (\pT) spectra and related/derived observables such as multiplicity, $\langle \pT \rangle$ and ratios, but also the observed ``ridge'' and the questions of quark-gluon plasma in pp collisions will be discussed. The goal of the review is on one hand to introduce the topics and provide references for scientists joining the LHC program while at the same time highlighting what we consider the most interesting results and open questions to inspire novel measurements.
\end{abstract}

\maketitle

\tableofcontents

\section{Introduction}

The LHC is, since it became operational, the energy frontier of particle physics. It is a diverse laboratory where:
\begin{itemize}
\item the standard model is being tested to impressive precision.
\item the search for physics beyond the standard model is ongoing in numerous final states.
\item the microscopic phenomenology of the quark-gluon plasma is being tested. 
\end{itemize}
A common denominator of these activities is that they all depend on our understanding of the microscopic laboratory itself: the proton-proton (pp) collision. The protons are build up of quarks and gluons, and their partonic interactions are described by quantum chromodynamics (QCD). For inelastic collisions, one can have diffractive collisions, where no color charge is exchanged, or non-diffractive collisions, where color charge is exchanged. Here we will primarily focus on the non-diffractive collisions.  In pp collisions, at the high energy of the LHC, several microscopic interactions will usually occur in each collision and each of these subcollisions can be characterized by the four-momentum transfer, $Q^2$. When $Q^2$ is large we call the interaction hard and we can typically utilize perturbative QCD to describe it~\cite{Campbell:2006wx}. However, when the $Q^2$ is low and the interaction is soft we reach the non-perturbative regime of QCD where we do not know how to calculate observables rigorously from first principles and we instead have to rely on phenomenological models and data-driven scaling relations to interpret the data. 
This review focuses on the same soft QCD in two regimes: 1) when all subcollisions are soft and the collision itself is therefore soft. This turns out to be the case for the bulk of pp collisions at LHC. 2) the soft subcollisions associated with one or more hard subcollisions, which is denoted the underlying event (UE).

Soft QCD is, like the LHC itself, a diverse research area. For some researchers at LHC, the soft QCD is an ``annoying'' background to their hard signal and the goal of tuning phenomenological models is to describe this background as precise as possible. For others, the soft QCD is the signal and it has turned out that the LHC is a discovery laboratory, which we will try to highlight in the later sections of this review. When it comes to theory, soft QCD seems to find itself in a difficult situation: to be able to compare precisely with data one essentially has to use full pp generator models. However, most people working on understanding details of the soft pp collisions from first principles are often unwilling or do not have the resources to include their developments in these ``simpler'' models. This is hopefully something that will change in the coming years because as we hope to show, the field of soft QCD has never been more exciting than it is now and there will likely not be a better opportunity to study soft QCD in the next two or more decades.

Let us now try to explain what we mean with soft QCD in more detail. As perturbative QCD only describes quarks and gluons while hadrons are measured, there is always non-perturbative QCD involved in any QCD calculation, but this is not what we are particularly focused on here. Here we are rather interested in the physics dominated by the soft subcollisions. What scale $Q^2$ is then soft? One would naively expect that for $Q^2 \gg \LQCD^2$, where $\LQCD \approx 200\,\text{MeV}$, the subcollision could be described by hard QCD, e.g., as a $2 \rightarrow 2$ partonic scattering. This would suggest that when we have particles with transverse momentum $\pT \gg \LQCD$ then they are the result of hard scatterings. However, this is not what we find experimentally. In fact, as we discuss in this review, there are many indications that the behavior of baryons, such as protons, remains distinctly different from hard QCD predictions out to $\pT \approx \gevc{8-10}$. This is a big surprise of LHC and one that is still not fully understood. So the soft scale is not set in stone but it seems that most particles with $\pT \leq \gevc{1-2}$ are produced in soft processes, while one has to go to $\pT > \gevc{10}$ to be in the fully hard regime, and so these are good experimental numbers to have in mind.

In this concise review paper, we aim to provide both an accessible introduction to soft QCD, supplemented with references for further reading, and a discussion of an eclectic selection of results and open questions that we find particularly intriguing. However, the expansive scope of soft QCD inevitably means that not all relevant topics could be addressed in detail. The structure of the paper is as follows: Section~\ref{sec:brief} offers a brief historical introduction to soft QCD physics. Section~\ref{sec:modelling} presents an overview of contemporary models employed in physics analyses. Section~\ref{sec:MPIs} provides details on the methodology for studying multi-parton interactions and evaluates the performance of UE tunes in Monte Carlo event generators. Section~\ref{sec:PID} examines the production of identified particles in greater depth, while Section~\ref{sec:collectivity} focuses on the phenomenon of collectivity. Finally, Section~\ref{sec:outlook} concludes with a discussion of future prospects in the field.

\section{A Brief Historical Introduction}
\label{sec:brief}

Soft QCD has been studied from early experiments at fixed-target facilities to
modern high-energy colliders. The measurement of cross sections, as well as multiplicity distributions and longitudinal and  transverse momentum (\pT) spectra gained most of
the interest, see for example
Ref.~\cite{Grosse-Oetringhaus:2009eis, Kittel:2005fu} for a pre-LHC overview.

Before the advent of QCD, Regge theory \cite{Collins:1977jy} offered a framework to discuss hadronic cross sections.  Central to this model is the Pomeron, which effectively captures the exchange of vacuum quantum numbers in soft QCD interactions.  At high energy, the total hadronic cross section is dominated by Pomeron exchange and is predicted to rise with centre-of-mass energy as $s^\epsilon$, where $\epsilon \approx 0.08$ is a small positive exponent.  This growth of the cross section can be related to the $x^{-\lambda}$ increase of parton densities in the proton towards small Bjorken-$x$ values \cite{Ellis:1996mzs}, with $x \propto \frac{1}{\sqrt{s}}$ and $\lambda$ a function that depends on $Q^2$.

\begin{figure}[htbp]
\includegraphics[width=\textwidth]{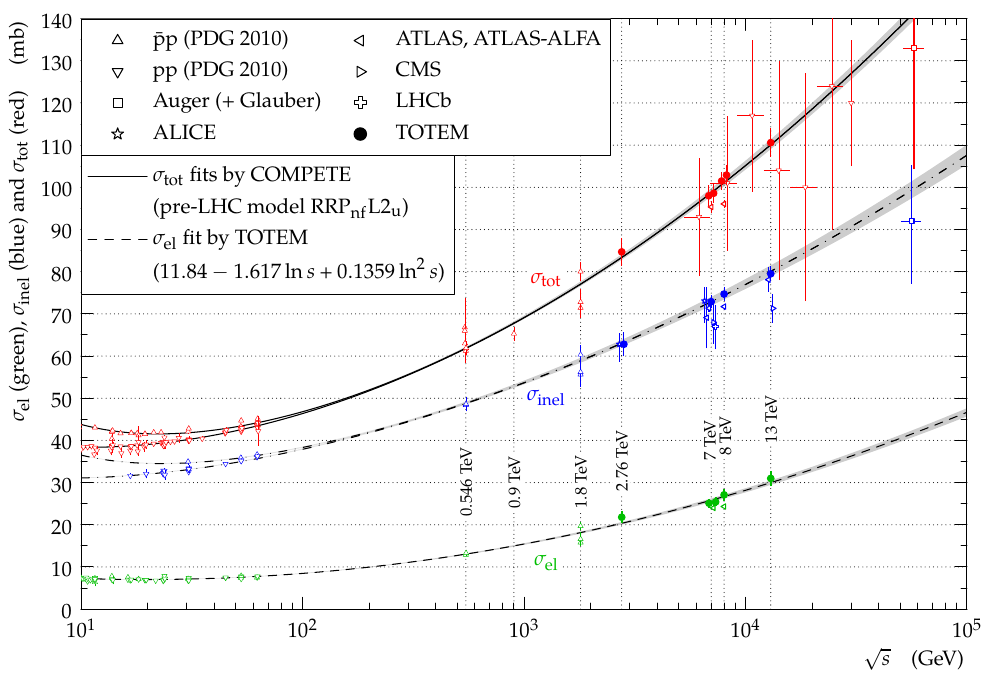}
\caption{Overview of elastic, inelastic, and total cross sections for pp  and ${\rm p\bar{p}}$ collisions as a function of center-of-mass energy, $\sqrt{s}$ (taken from \cite{TOTEM:2017asr}).}
\label{fig:ppcrosssection}
\end{figure}

This simple power-law rise would eventually conflict with the Froissart-Martin unitarity bound \cite{Froissart:1961ux}, which constrains the growth of  cross sections to a $\ln^2 s$ dependence. Even though the cross sections measured at the LHC remain significantly below this limit (numerically still a factor $\sim 100$ below this bound \cite{Martin:2008xb}), the ratio of the elastic to total cross section ($\approx 0.3$ at LHC energies) increases with $\sqrt{s}$ and is approaching the theoretical maximum of $\sigma_{\rm el}/\sigma_{tot} = 0.5$ for a black disk.  To account for the observed energy dependence, the exchange of multiple Pomerons must be considered.  In terms of parton densities this could be connected to non-linear effects occurring below the saturation scale $Q_s$ \cite{Kovchegov:1999yj}.  This scale increases with center-of-mass energy and reaches  $\sim \gevc{1-2}$ for protons at the LHC, and higher for heavy ions.  Experimental results from the LHC and other high-energy colliders confirm that total and elastic cross sections are in agreement with a log-square rise (see \textbf{Figure \ref{fig:ppcrosssection}}).

Particle production in inelastic collisions is linked to the elastic cross section via the optical theorem, which connects the imaginary part of the forward scattering amplitude to the total cross-section. In Regge theory, this connection introduces cut Pomeron-exchange diagrams, which result in multiple particle production chains \cite{Abramovsky:1973fm}. Energy-momentum conservation dictates that a hadronic system with invariant mass $M$ will fragment into final-state particles with rapidities ranging between $\pm \ln(M/m)$ in the center-of-mass system, where $m$ is the mass of the lightest final-state particles, typically pions.  The full rapidity distribution will thus be described by a superposition of rapidity distributions of individual production chains with different invariant masses, with maximal rapidity given by $\pm \ln(\sqrt{s}/m_{\rm p}) \approx \pm 9.6$ for LHC energies of $\sqrt{s} = 13\ {\rm TeV}$. This framework is crucial for understanding multi-particle production in high-energy hadron collisions, as it illustrates the strong connection between the saturation of cross sections and parton densities on one side, and the emergence of multiple particle production systems on the other.

The multiplicity distribution gives the probability
that a certain multiplicity is observed in a pp collision and is only
measurable for charged particles. The observed particle multiplicity
distribution follows approximately a negative binomial distribution,
which captures the overdispersion seen in the data (i.e., the variance of the
distribution is larger than the mean). In many cases, the multiplicity
distributions exhibit long tails, indicating rare events with exceptionally
high particle production.

\begin{figure}[htbp]
\begin{minipage}{0.5\textwidth}\centering
\includegraphics[width=\linewidth]{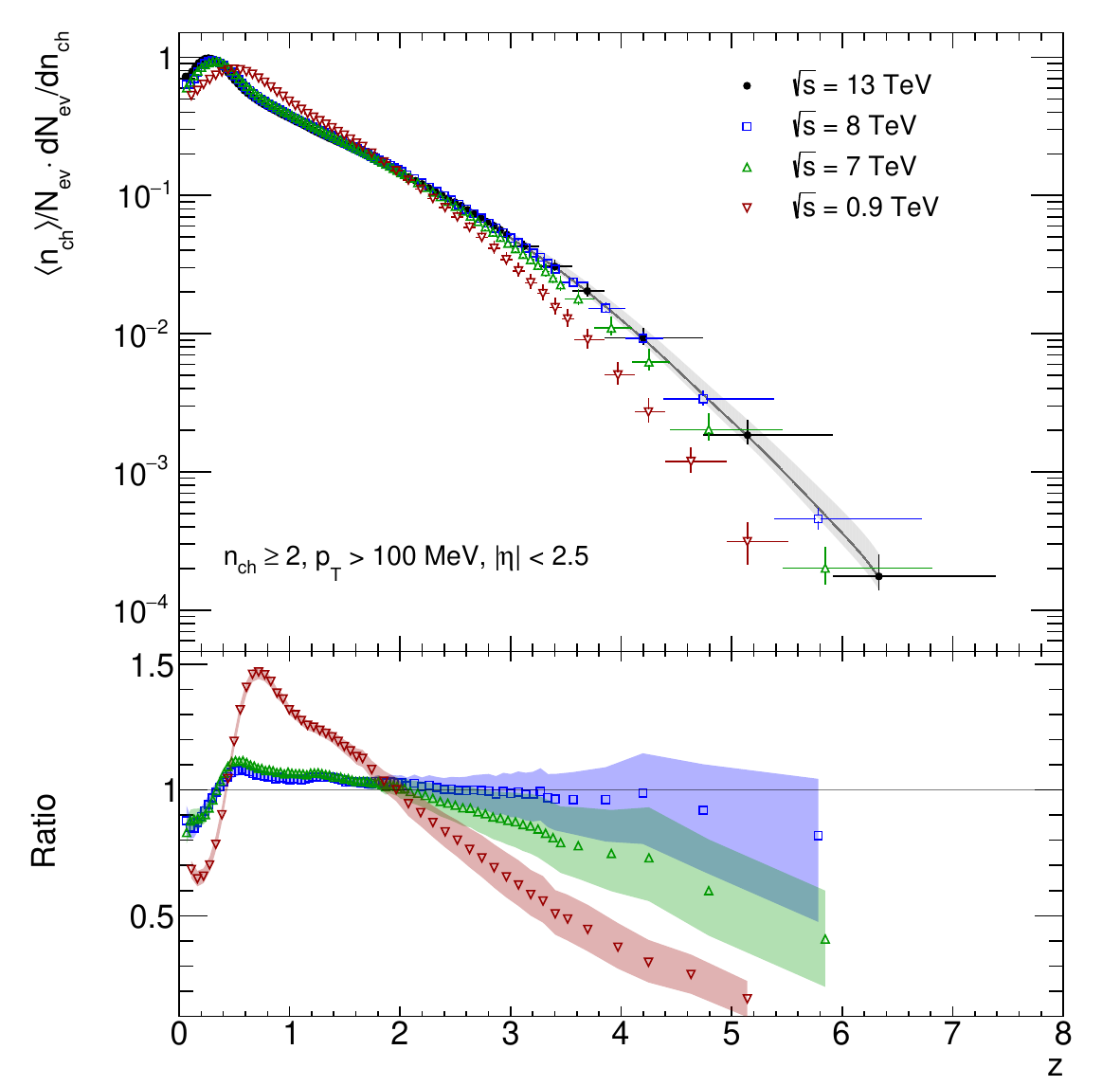}
\end{minipage}
\begin{minipage}{0.5\textwidth}\centering\vspace{5mm}
\includegraphics[width=0.85\linewidth]{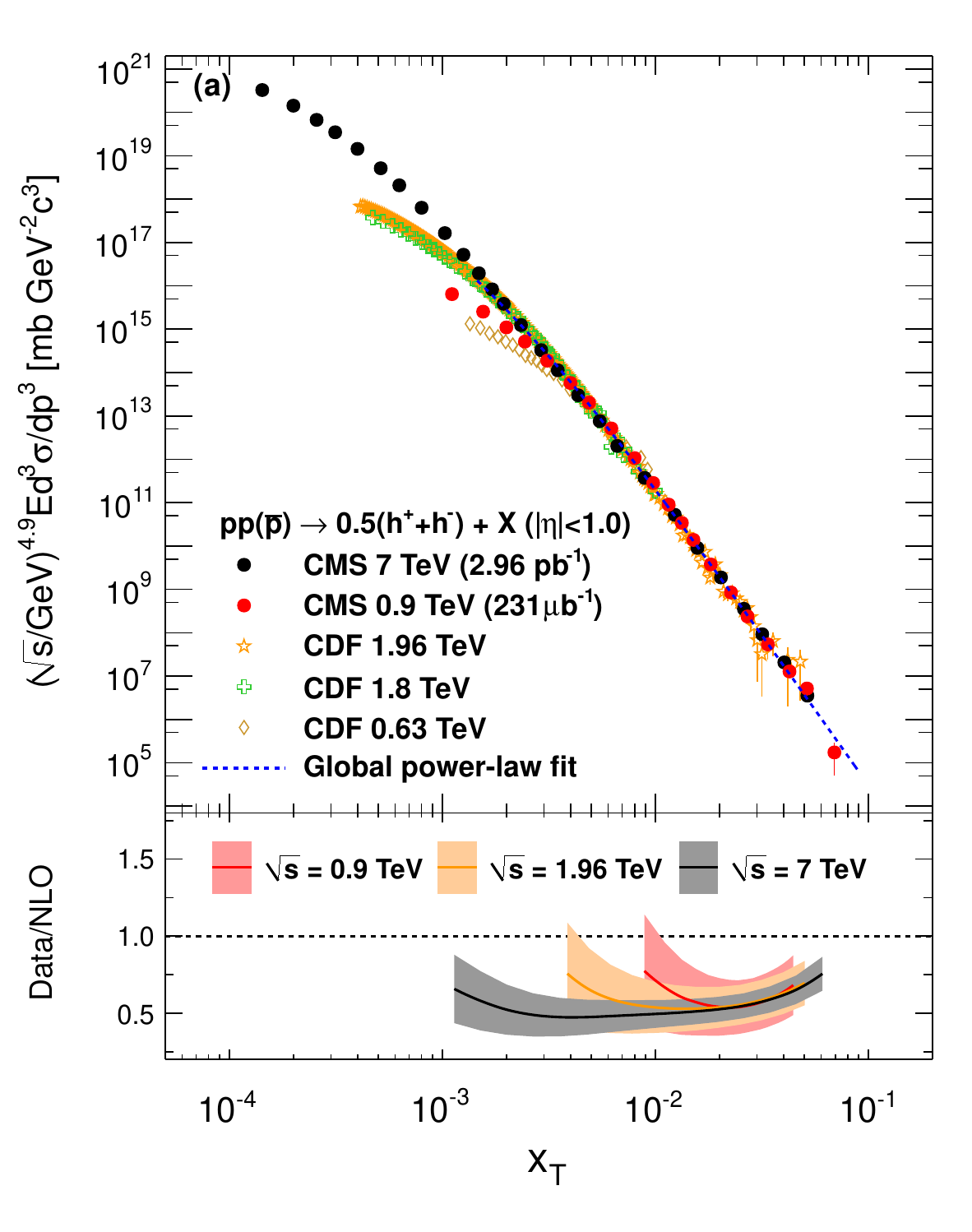}
\end{minipage}
\caption{(left) KNO scaled primary charged-particle multiplicity distributions from the ATLAS experiment as a function of the scaled multiplicity for various centre-of-mass energies, together with their ratio to the $\sqrt{s} = 13$ TeV distribution (taken from \cite{Kulchitsky:2022gkm}); \\
(right) Inclusive charged particle invariant differential cross section from the CMS and CDF experiment as a function of the scaling parameter \xT for various centre-of-mass energies, together with their ratio to NLO calculations (taken from \cite{CMS:2011mry}).}
\label{fig:knoscaling}
\end{figure}

In the early days of soft QCD studies, Koba-Nielsen-Olesen (KNO) scaling \cite{Koba:1972ng} was observed in particle multiplicity distributions. KNO scaling posits that the normalized multiplicity distributions should collapse into a single universal curve when scaled by the mean multiplicity, implying that particle production has a universal behavior across different energies. This phenomenon was initially supported by experimental data from lower-energy collisions. However, as measurements at higher energies were conducted, particularly at the LHC, a significant violation of KNO scaling was observed, as illustrated in \textbf{Figure \ref{fig:knoscaling}}(left).
The superposition of multiple distinct topologies with varying average multiplicities broadens the overall multiplicity distribution. Consequently, the breakdown of KNO scaling suggests that particle production cannot be explained by a single component or mechanism but instead points to a multi-component structure in the final state. 

The \pT-spectra describe the average distribution of particles as a function
of \pT. The spectra are typically normalized as an invariant yield, e.g.,
$1/(2\pi\pT)d^2N/(dyd\pT)$, where $y$ is the rapidity, and sometimes they are
normalized to the cross section, $\sigma$, instead of per event. The \pT
spectra typically display a power-law behavior at high \pT and an exponential
drop at low \pT. This behavior reflects the different underlying physical
processes: soft interactions dominate at low \pT, while harder QCD processes
contribute the tail at higher transverse momenta. Indeed, the large \pT behaviour can be captured by the concept of \xT scaling \cite{Blankenbecler:1972cd}. Here, \(\xT = 2\pT/\sqrt{s}\) is a dimensionless variable that scales the transverse momentum \pT by the center-of-mass energy \(\sqrt{s}\). In hard scattering processes, it is expected that the spectra measured at different collision energies should collapse onto a universal curve when plotted as a function of \xT. This scaling behavior reflects the self-similar nature of particle production processes. The violation of \xT scaling in the soft region indicates that purely perturbative approaches are insufficient to describe the entirety of the spectrum, and non-perturbative effects and multiple scattering processes must be included for a complete picture of particle production.  Interestingly, the absolute \pT threshold below which non-perturbative effects are important appears to increase with centre-of-mass energy, reaching values above $\sim \gevc{10}$ at the LHC, as can be seen in \textbf{Figure \ref{fig:knoscaling}} (right).

In summary, even before QCD, Regge theory had introduced the idea of producing multiple particle chains. Modern Monte Carlo event generators therefore incorporate the concept of multiple parton interactions (MPI), where multiple partons within colliding protons can interact independently. 

The LHC provides a unique opportunity to study soft QCD.  At higher center-of-mass energies, the importance of MPIs increases, significantly contributing to the particle multiplicity. MPIs thus account for both the rapid increase in average multiplicity and the greater variability in charged particle counts observed at the LHC and other high-energy colliders.

\section{Phenomenological Modeling Perspectives}
\label{sec:modelling}

As we have seen in the last two sections, some features of soft QCD results
can be understood in terms of QCD-inspired scaling relations and/or
fits. However, to go towards  quantitative descriptions of the measurements one needs phenomenological
models that generate full events. The most popular phenomenological models
used to describe pp collisions at LHC have been explained in detail in
Ref.~\cite{Buckley:2011ms}. Here, we will give a brief introduction to the
main basic elements of soft QCD modeling only, to highlight the major
differences between models and setup the discussion in the later sections.

Monte Carlo event generators incorporate detailed models to simulate soft pp
collisions and the underlying event. At the partonic level, the most important ingredient in the
models are MPIs as they primarily determine the
activity and structure of the event~\cite{Sjostrand:1987su}. Here and in the following we always refer
to the soft QCD part
  of the event. Possible hard scatterings will of course also affect the activity and
  structure but for the soft QCD observables we are interested in in this
  review, we typically can get around this issue, e.g., by looking in the
  transverse region, see Sec.~\ref{sec:UEobs}. To accurately describe data one also has
to include possible coherence effects between different interactions, e.g.,
color reconnection (CR)~\cite{Sjostrand:1987su}. The final important model ingredient that we will
discuss is hadronization
where it is also important to handle the color connection to beam remnants.


The models we will mainly focus on here are \pythia\cite{Bierlich:2022pfr}, \herwig~\cite{Bellm:2015jjp}, and \epos~\cite{Pierog:2013ria,Werner:2023zvo}. The first two generators are the most popular pp generators at LHC, which includes a detailed handling of soft QCD, and they are systematically compared to LHC
data at the \mcplots website \url{https://mcplots.cern.ch/} \cite{Karneyeu:2013aha}. The final generator, \epos,
is different from the other generators in that it also contains the partial
formation of a quark-gluon plasma (QGP) in pp collisions.

\subsection{Multiple-Parton Interactions}

Let us first discuss this in a more abstract sense that can apply to all
models. MPIs are in all cases treated as a sum of incoherent subcollisions,
typically $2 \rightarrow 2$ partonic interactions. The QCD cross sections for these individual subcollisions diverge as \(\pT \to 0\), ultimately exceeding the total cross section. This  is not inherently problematic, as the ratio of the partonic QCD cross section to the total cross section can be interpreted as representing the number of MPIs occurring. Nonetheless, to prevent the cross sections
from becoming unphysically large at low $\pT$, they must be
regularized. This regularization can be linked to color screening, which limits the resolution of individual partons, as implemented, for instance, in \pythia, or to initial-state saturation effects, as done by \epos.

Let us focus on a concrete example. In \pythia, subcollisions are modelled as
$2 \rightarrow 2$ partonic processes and a transverse momentum cut-off
scale, $\pTzero$, is introduced to regularize the cross section:
\begin{equation}
  \frac{{\rm d}\hat{\sigma}}{{\rm d}\pT^2} \propto \frac{\alpha_S^2(\pT^2)}{\pT^4}\to \frac{\alpha_S^2(\pTzero^2+\pT^2)}{(\pTzero^2 + \pT^2)^2},
\end{equation}
where $\pTzero$ depends on the collision energy:
\begin{equation}
    \pTzero(s) = \pT^{\rm ref}\left(\frac{\sqrt{s}}{\sqrt{s_0}}\right)^\epsilon,
\label{eq:mpi_pt0}
\end{equation}
with $p_{\rm T}^{\rm ref}$ and $\epsilon$ as tuneable parameters. One finds
that \pTzero increases with energy, which is expected due to the increase of
low-$x$ partons in the proton wavefunction, and is of order \gevc{2} at LHC,
i.e., we are only sensitive to length scales smaller than 0.1\,fm, which is much smaller than the proton radius. Note that in a saturation picture, \pTzero is related to $Q_s$. 
Additionally, an impact parameter dependence is introduced, making the number
of MPI reliant on the spatial overlap of the protons' transverse profiles so
that central collisions produce more MPIs than peripheral ones.

\subsection{Intrinsic transverse momentum and initial and final state radiation}

Intrinsic transverse momentum (\kT) represents the perpendicular momentum component of partons within nucleons and plays a critical role in modeling non-perturbative QCD effects. In processes like Drell-Yan production, intrinsic \kT influences the low \pT spectrum of dileptons, while initial state radiation (ISR), which describes gluon emissions by partons before hard scattering, introduces perturbative QCD corrections at higher \pT.   Intrinsic \kT values on the order of \gevc{1-2}, increasing with center-of-mass energy, were determined through an analysis of Drell-Yan data  \cite{CMS:2024goo}.  

Final state radiation (FSR), involving gluon emissions from outgoing partons post-scattering, further redistributes momenta. Together, intrinsic \kT, ISR, and FSR provide a comprehensive framework for simulating angular and momentum correlations in the final state.  A notable example is jet production in high-energy collisions, where these effects contribute to deviations in azimuthal correlations, particularly in back-to-back jet configurations \cite{CMS:2019joc}. 

Monte Carlo event generators model ISR and FSR as a parton shower, where successive emissions are computed based on the Altarelli-Parisi splitting functions that describe the probability of a parton splitting into a pair of partons. In ISR, each emission reduces the parton’s energy and modifies its transverse momentum, leading to a cascade of radiation that results in a more complex initial state. FSR accounts for the fragmentation of outgoing partons into hadrons, forming jets and shaping the final event topology.

\pythia employs an interleaved approach to handle MPI alongside ISR and FSR. In this method, MPI, ISR, and FSR steps are dynamically combined in a common ordering sequence. This approach enables \pythia to account for correlations and competition between these interactions, leading to a more accurate depiction of the particle production process.

\subsection{Hadronization and color reconnection}
\label{sec:hadronization}

Once the MPI model has generated a colored final state, it has to be
hadronized. This is an area with substantial differences between
generators. For example, \pythia uses the Lund string model and \herwig uses a
cluster model. An important assumption of most hadronization models is ``Jet
Universality'', which means that the final objects that hadronize, e.g.,
strings or clusters, and the parameters used in the hadronization of these
objects are universal. This implies that the parameters can be tuned using
better understood data from $\ee \rightarrow \qqbar$ so that their parameters
are fixed for pp collisions. Finally, each generator also have to implement a
scheme for how it handles the beam remnants in the hadronization process.

To accurately describe data it turns out that one has to add a step in between
MPIs and hadronization where one breaks the incoherence of the MPIs. This is
very important for describing the observed rise of \meanpT with multiplicity
in pp collisions, as will be explained next. The multiplicity grows with the
number of MPIs. If MPIs were incoherent, then they should on the average
produce the same \meanpT because of Jet Universality and so there would be no rise of \meanpT with
multiplicity. \pythia, for example, has a color reconnection (CR) model that
adjusts the color flow between partons originating from different scatterings
to minimize the total string length. This allows \pythia to describe the
\meanpT increase. This breaking of incoherence is critical for describing the data and
it turns out that even more advanced models are needed to describe the
strangeness enhancement discussed in Sec.~\ref{sec:pid:spectra}.  Surprisingly, CR is also relevant to $W^+W^-$ pair production in $e^+e^-$ annihilation when the $W$ bosons decay hadronically, and influences the systematic uncertainties in the measurement of the $W$ mass.  We will
return to the more advanced CR schemes later in the text when we discuss the
data that requires the models to introduce them.

The most extreme way to break the incoherence is found for \epos. \epos assumes
that a QGP will form in the dense regions of a pp collision (those regions
with many MPIs). The MPI subcollisions are therefore classified into two distinct groups: a
non-QGP ``corona'', which hadronize more or less independently, while
subcollisions in a dense QGP region forms a common ``core'' that expands hydrodynamically
and cools before hadronizing according to the statistical thermal
model~\footnote{The statistical thermal model has been hugely successful in
  describing integrated particle yields in AA collisions, see for example
  Ref.~\cite{Andronic:2017pug}.}. In \epos, the rise of
\meanpT with multiplicity occurs because the importance of the core grows with
multiplicity and because the expansion of the core will give rise to a radial
flow that boosts the \pT. For example, assuming a hadron is produced with zero
\pT in the QGP rest frame, it will end up with the following \pT in the laboratory
frame:
\begin{equation}
  \label{eq:boost}
  \pT = \gamma \beta_{\rm r} m, 
\end{equation}
where $\beta_{\rm r}$ is the radial flow velocity and $m$ is the mass of the
particle. As can be seen, this will introduce a mass dependence for the
\meanpT growth with multiplicity, which is also observed in data, see for
example~\cite{CMS:2012xvn}. Importantly, it can be shown that the CR of \pythia
effectively also produce a boost like Eq.~\ref{eq:boost} and therefore also
produce this mass dependence~\cite{OrtizVelasquez:2013ofg}. This opens a big
question of to what degree CR and hydrodynamics is at work in pp
collisions. Is it for example all CR and there is no QGP or is it all
hydrodynamics and there is no CR? This question is still not resolved and is
probably the biggest open questions when it comes to soft QCD at the LHC.

\subsection{Summary of Generators}

\begin{table}[h]
\tabcolsep7.5pt
\caption{Summary of the main differences between the pp generators covered in this review. The structure of the table follows the structure of the previous subsections.}
\label{tab:generators}
\begin{center}
\begin{tabular}{@{}l|c|c|c|c@{}}
  \hline
  Generator & Ref. & Soft MPIs & CR or similar & Hadronization \\
  \hline
  \pythia 8 & \cite{Bierlich:2022pfr}  & $2 \rightarrow 2$ partonic
           & CR, Ropes, Junctions & Lund strings \\
  \herwig 7 & \cite{Bellm:2015jjp} & $2 \rightarrow 2$ partonic & Statistical CR, baryonic R & Clusters \\
  \eposlhc & \cite{Pierog:2013ria} & Pomeron exchanges
           & Core-corona model & Thermal core \\
  \hline
\end{tabular}
\end{center}
\end{table}

\textbf{Table \ref{tab:generators}} summarizes the features of the generators
that were discussed previously in this section. For \epos, the relatively old \eposlhc version is listed while there is also a more
modern \epos 4 version~\cite{Werner:2023zvo}. However, currently \eposlhc is
the most used version at LHC. \\ The main differences between the generators are for the CR and hadronization, which are often interleaved: the CR implemented in \pythia is handled at the string level while the core-corona formation in \epos controls the hadronization. Even CR concretely leads to modified hadronization in the models, the physics interpretation is not clear and one can also think of CR as a modification of the MPIs, e.g., to mimic saturation. 

For modeling of MPIs there are alternative theoretical frameworks such as the
Color Glass Condensate (CGC), which is derived from first principles, see
\cite{Gelis:2010nm} for an overview. The CGC takes advantage of the growth of
low-$x$ gluons in the proton wavefunction with the center-of-mass energy. Because
of these dense gluon fields at LHC, the MPIs are in the CGC framework modeled
as collisions between classical fields and the regularization enters via the
saturation scale, $Q_s$. As the LHC is the energy frontier for the foreseeable
future, it would be interesting to understand better if such a framework can
be incorporated into pp generators and what the observables that would be
most affected at the LHC would be and/or if there would be unique predictions of
such a framework.

\section{Multiparton Interactions and Underlying Event Tunes}
\label{sec:MPIs}

Inspired by Regge-Gribov theory, which allows for events with multiple cut Pomerons--each producing a sequence of low-$p_T$ hadrons—and motivated by experimental observations, Monte Carlo event generators incorporate detailed models to simulate the underlying event. These models account for contributions from MPI, intrinsic \kT, ISR and FSR, color reconnection, and beam-beam remnants (BBR). The models introduce several free parameters, which are tuned to minimum bias and UE data by experimental collaborations. Specific observables have been designed to be sensitive to UE activity. Current tunes successfully describe the underlying event across various final states, including multijet, Drell-Yan, Z and W boson production, and top quark pair production.  An excellent review of MPI at the LHC can be found in \cite{Bartalini:2018qje}.

\subsection{Underlying Event Observables}
\label{sec:UEobs}

A widely used approach \cite{CDF:2001onq} for systematically measuring the underlying event in pp collisions relies on identifying a leading object in the event, which could be the charged particle or jet with the highest transverse momentum. This leading object is associated with the hard scattering of the event. The transverse momentum of the leading object, $p_T^{\rm leading}$, is used as a reference for UE measurements.  

The event is divided into different regions of azimuthal angle $\phi$ relative to the leading object’s direction: 
\begin{itemize}
    \item The {\em toward} region is defined as the direction of the leading object ($|\Delta\phi| \le 60^\circ$), where contributions mainly come from the hard scatter.
    \item The {\em away} region is located in the direction opposite to the leading object
($|\Delta\phi| > 120^\circ$), and typically contains the recoil from the hard scatter.
\item The {\em transverse} regions are defined as the regions perpendicular to the leading object ($60^\circ < |\Delta\phi| \le 120^\circ$), this is where the UE activity is expected to dominate. Since it is far from both the hard scatter and its recoil, it primarily contains contributions from MPI, ISR, FSR, and BBR. 
\item To further refine the measurement, the transverse region can be split into two subregions: {\em transMIN} is the transverse region with the minimum activity and is particularly sensitive to MPI and BBR since it minimizes the contribution from ISR or FSR, and {\em transMAX} which is the transverse region with the maximum activity which may include ISR and FSR in addition to MPI and BBR. By comparing transMIN and transMAX, the method can distinguish between ISR/FSR and MPI contributions -- e.g. by computing the difference of transMAX and transMIN, one can single out the contribution from ISR/FSR.
\end{itemize} 

Two main observables are studied to characterize the underlying event in these regions:  multiplicity $N_{\rm ch}$ and the scalar sum of transverse momentum $\sum p_T$ of the charged particles.  Although multiplicity is not an infrared-safe observable, it is closely linked to the number of MPIs. In contrast, the sum of transverse momentum is primarily governed by the overall energy flow distribution w.r.t.\@ the leading object.  These quantities are then measured as a function of $p_T^{\rm leading}$.

\subsection{Monte Carlo Tunes}

UE tunes are derived using tools such as \professor \cite{Buckley:2009bj} and \rivet \cite{Bierlich:2019rhm}. These tools enable experimental collaborations to compare simulated event samples generated by Monte Carlo event generators, like \pythia and \herwig, with real data. \professor is used for automated tuning: it parametrizes the MC generator's predictions and fits them to minimize the difference between the simulation and experimental data. \rivet (and \mcplots) provides a framework for comparing distributions of observables sensitive to the underlying event, such as charged-particle multiplicities and transverse momentum sums, between data and simulation. This combination allows experiments to derive the best-fit parameters for MC generators, resulting in "tunes" that accurately describe the underlying event across a range of processes and energies. Through this process,  different tunes, that significantly improve the reliability of simulated collisions at the LHC, have been obtained by various authors. In the next section we compare  observables to the following specific tunes obtained by ATLAS and CMS:

\begin{itemize}

\item A significant contribution is the CP5 tune \cite{CMS:2019csb} for \pythia 8, which was created by CMS using the parton distribution functions (PDFs) from NNPDF 3.1. The CP5 tune is a multipurpose tune, aiming for a consistent description of underlying and minimum bias observables at several collision energies and a reliable prediction of the UE simulation in various processes when merged with higher-order ME calculations. The tune was extensively validated and showed that using NNLO PDFs significantly improves the modeling of both soft and hard processes in proton-proton collisions, making CP5 one of the most reliable tunes for high-precision LHC measurements.

\item One of ATLAS'  widely used tunes is the A14 tune \cite{TheATLAScollaboration:2014rfk} for \pythia 8, which was developed based on a combination of  observables from both minimum-bias events and Drell-Yan processes. The A14 tune uses the NNPDF 2.3 LO PDF set. It was optimized using data from minimum-bias and UE observables collected at a center-of-mass energy of 7 TeV. The tune was validated against a wide variety of processes, including Z-boson production and jet production, demonstrating good agreement with experimental data. 

\end{itemize}

\subsection{Tune validation}

\begin{figure}[htbp]
\begin{minipage}{0.5\textwidth}\centering
\includegraphics[width=0.9\linewidth]{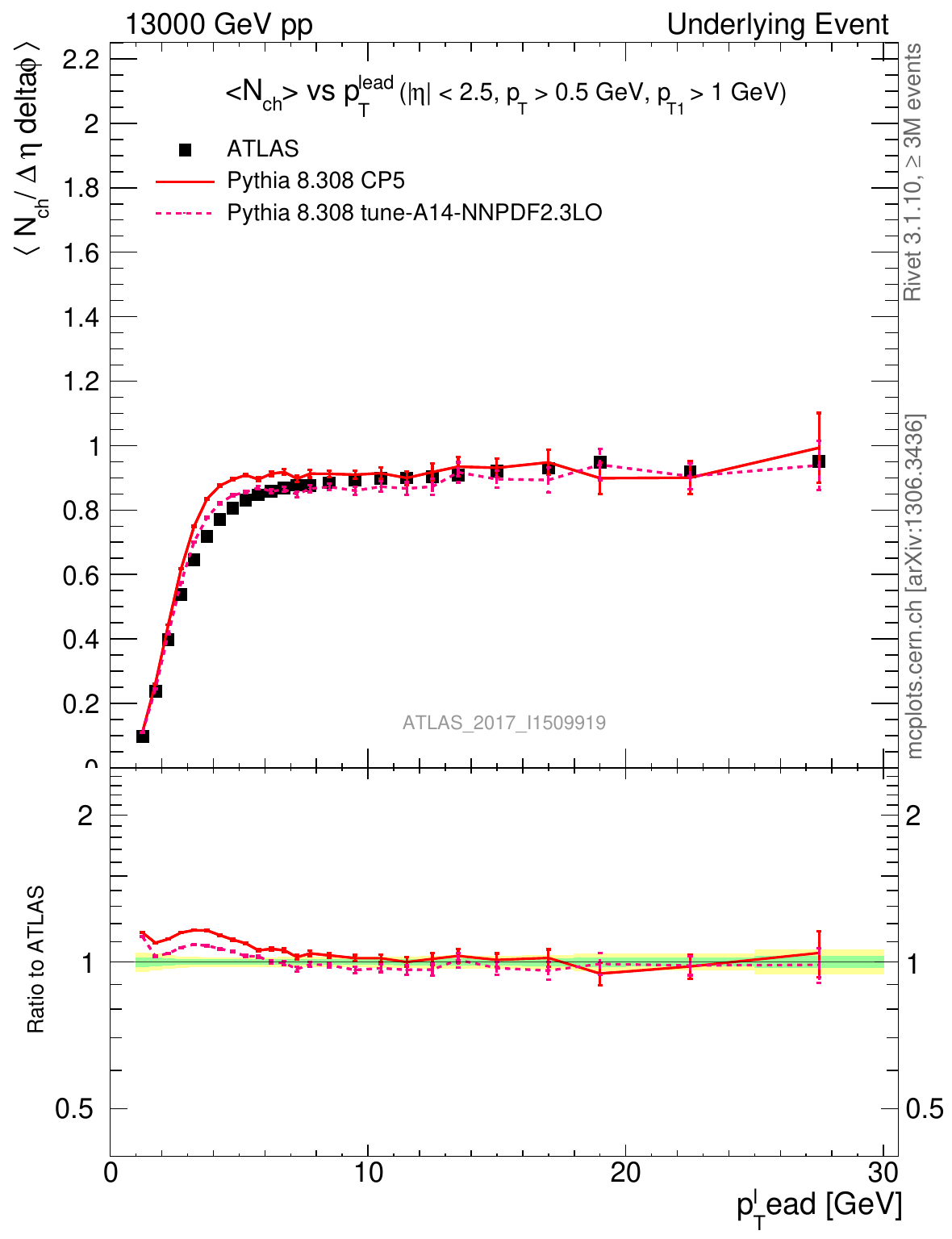}
\end{minipage}
\begin{minipage}{0.5\textwidth}\centering
\includegraphics[width=0.9\linewidth]{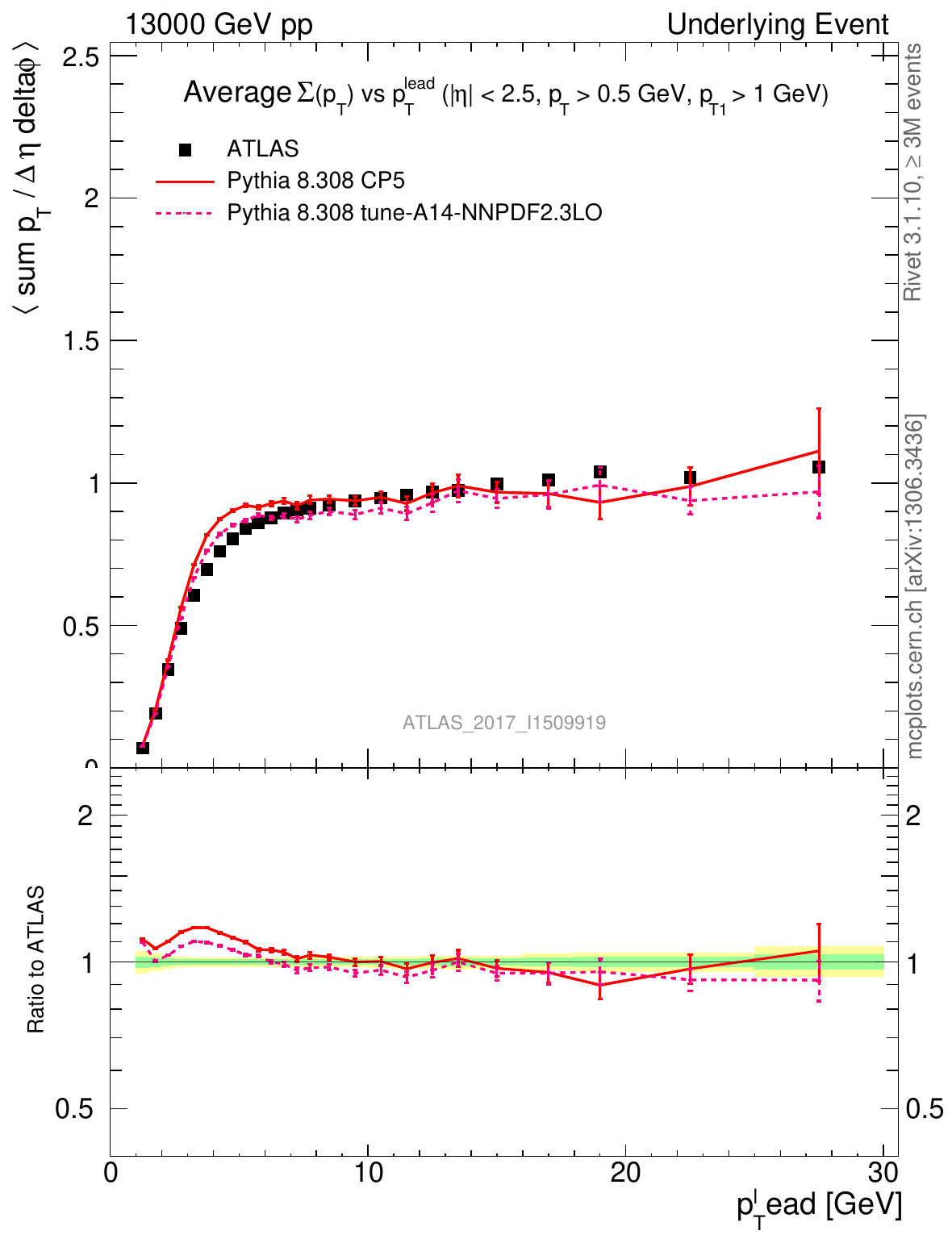}
\end{minipage}
\begin{minipage}{0.5\textwidth}\centering
\includegraphics[width=0.9\linewidth]{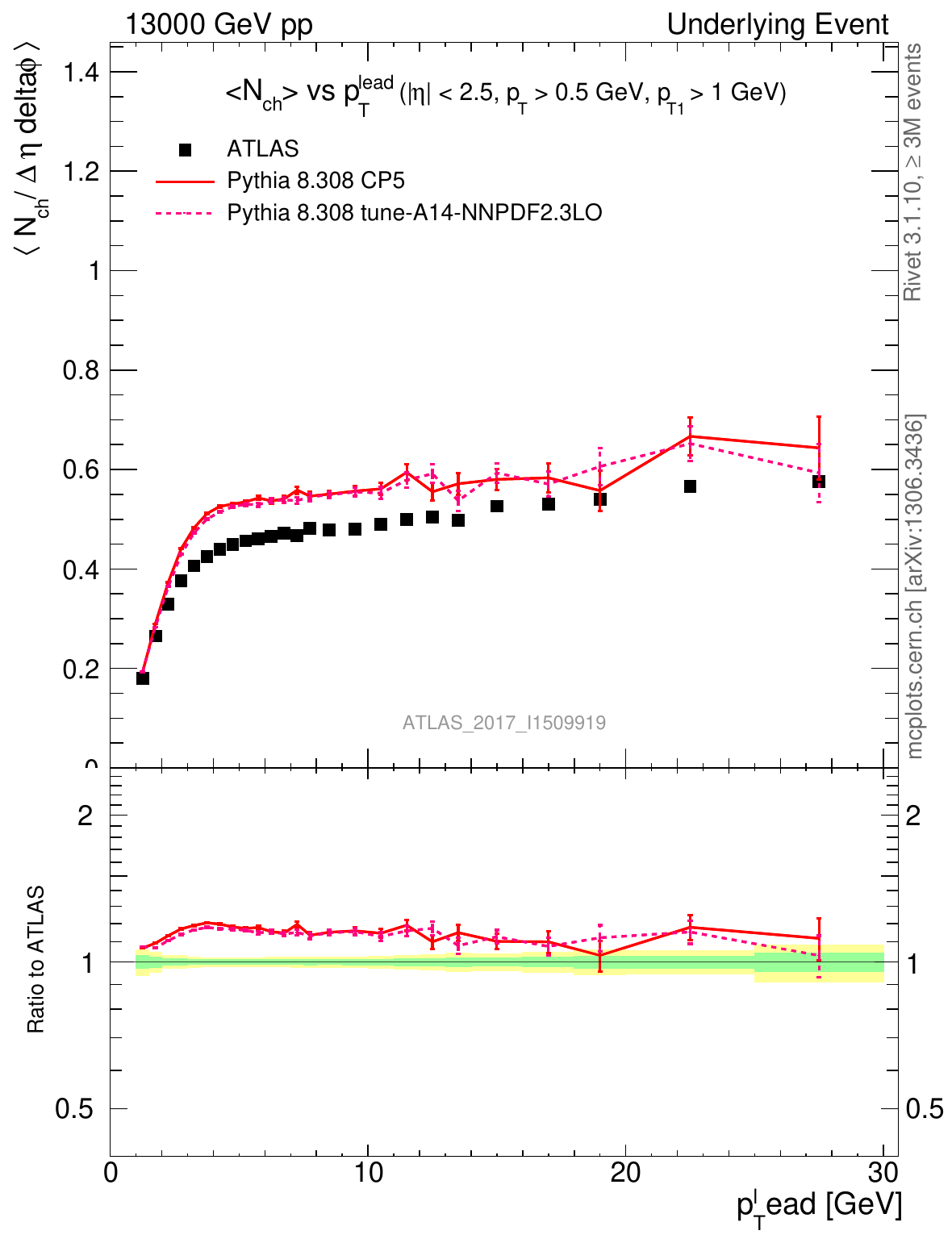}
\end{minipage}
\begin{minipage}{0.5\textwidth}\centering
\includegraphics[width=0.9\linewidth]{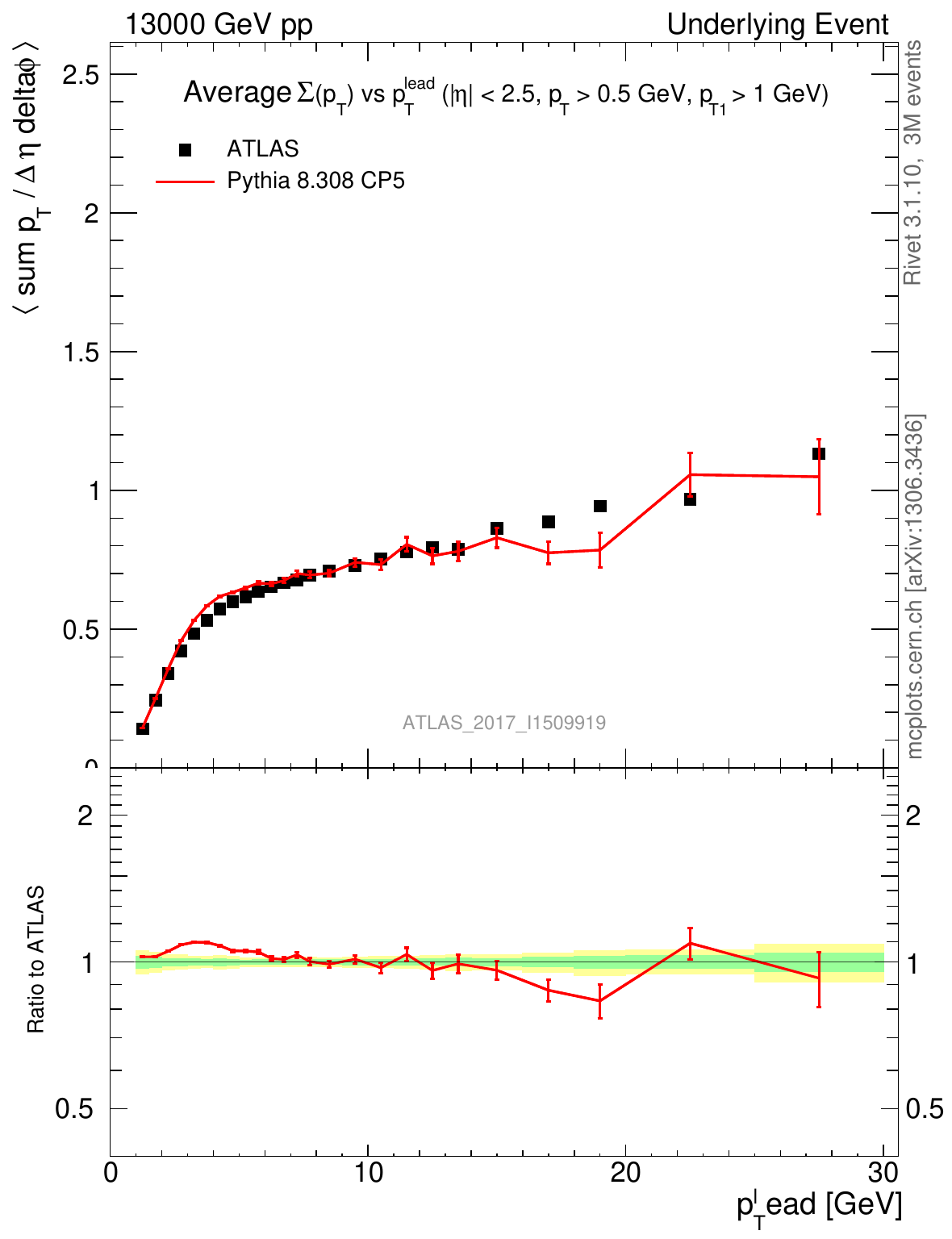}
\end{minipage}
\caption{UE observables are shown as functions of the transverse momentum of the leading charged particle. The top row presents the average charged-particle multiplicity (left) and transverse momentum sum (right) in the transMIN region. The bottom row displays the same observables for the transDIFF region. These plots were obtained using the \url{mcplots.cern.ch} tool.}
\label{fig:uetunes}
\end{figure}

\textbf{Figure \ref{fig:uetunes}}  demonstrates how different tunes reproduce various UE observables. The top row presents the average charged-particle multiplicity and \pT sum in the transMIN region as a function of the leading charged particle's \pT. This region is predominantly sensitive to MPI. Both the ATLAS and CMS tunes provide a reasonable description of the ATLAS data, although slight discrepancies appear around the turn-on region at $\pT^{\text{lead}} \sim \gevc{7}$. The bottom row shows these same distributions in the transDIFF region, which emphasizes initial-state and final-state radiation (ISR/FSR). While the tunes generally succeed in capturing the \pT sum, they tend to overestimate the charged-particle multiplicity. This outcome highlights the ongoing complexity of UE tuning and shows that certain tensions remain in accurately modeling these observables.

CMS has conducted detailed studies of color reconnection effects within the underlying event \cite{CMS:2022awf} and explored different models of color reconnection implemented in \pythia 8, particularly those based on MPI, QCD-inspired models, and gluon-move models. These studies compared the default CP5 tune (which uses the MPI-based CR model) with alternative CR models, and showed significant potential for improving the description of soft particle production and UE observables, providing a better fit to charged-particle multiplicity data and event shapes in the transverse region.

ATLAS has conducted UE studies focusing on strange hadron production \cite{ATLAS:2024nbm}, including $K_S^0$, $\Lambda$, and $\bar{\Lambda}$, which are sensitive to hadronization effects. These studies revealed that while the current MC models generally describe the data well, discrepancies remain in certain kinematic regions. For example, the observed yields of strange hadrons and their correlations with charged-particle multiplicity in the transverse region, used as a proxy for the number of MPIs, showed tensions with the predictions from the standard tunes. 
 
It is important to note that, beyond the specific UE observables discussed in Section \ref{sec:UEobs}, other properties of the final state are also affected by soft QCD. For example, measurements of transverse sphericity, as reported in \cite{ALICE:2012cor,ATLAS:2012uka}, show that sphericity increases with charged multiplicity, which correlates with the number of MPI. This effect is consistent with the expectation that additional MPI contribute to a less jet-like event topology. However, current tunes tend to underestimate this impact, predicting more jet-like final states than observed. This is confirmed in \cite{ALICE:2019dfi} where the average \pT of particles in jet-like events is overestimated by models, while the same observable is well described in events by more spherical final state topologies.

Rather than focusing solely on particle activity in the transverse plane, one can examine separations in rapidity. First measurements of the underlying event at forward rapidity at $\sqrt{s} = 0.9$, 2.36, and 7 TeV were reported in \cite{CMS:2013yli}. These studies were extended to $\sqrt{s} = 13$ TeV, for instance in \cite{CMS:2019kap,ALICE:2021poe}, providing valuable data for further validation of underlying event (UE) models.  

Similar arise findings from studies of forward-backward \cite{ATLAS:2012as} and long-range correlations in rapidity \cite{ATLAS:2016rbh}, which decrease with increasing charged multiplicity. Notably, these long-range correlations appear to be universal across different collision systems, including proton-proton, proton-lead, and lead-lead interactions.

Finally, it is worth noting that the tuning efforts discussed in this section are based on the assumption that MPIs are azimuthally isotropic, as this is how they are implemented in Monte Carlo simulations. However, there is no inherent reason to rule out the possibility of MPIs being correlated with a preferred azimuthal direction, such as the collision plane. As demonstrated in \cite{Alderweireldt:2012kt}, introducing such azimuthal correlations can reproduce the long-range, near-side correlations in proton-proton collisions initially observed by CMS \cite{CMS:2010ifv} (see Sec. \ref{sec:collectivity}).
 
\section{Particle Production of identified particles}
\label{sec:PID}

In this section we will focus on measurements of identified hadrons. As the
review is focused on soft particle production, we will limit ourselves to
hadrons containing up, down, and strange quarks and their associated
antiquarks. There are three additional complementary aspects of particle production that can
be studied with identified hadrons:
\begin{itemize}
\item Effects related to mass.
\item Effects related to quark content, e.g., strangeness.
\item Baryon vs.\ meson differences.
\end{itemize}
While the mass typically enters each model in a straightforward way, the
latter two provide fundamental insights into how quark masses and hadron type
affects particle production. For this reason they will be the focus
here. However, it should be pointed out that even mass and baryon vs.\
meson differences are well separated in models they are not so easy to separate
experimentally as baryons typically are heavier than mesons. This means that
some experimental observations can be explained in one model by a mass effect
and in another model by a baryon vs.\ meson effect.

Let us start by focusing on the biggest paradigm change that has resulted from
measurements of identified particle production at LHC. As we discussed in
Sec.~\ref{sec:hadronization}, the traditional pp paradigm with MPIs and Jet
Universality leads to pp collisions where the properties of subcollisions are
independent of multiplicity and constrained using data from
$\ee \rightarrow \qqbar$. For this reason, one expects that particle ratios of
strange hadrons to charged pions should not depend substantially on particle
multiplicity. It was therefore a big surprise when ALICE measured that the
\pT-integrated ratios of cascades to charged pions, $\XI(ssd)/\pi$ and
$\OM(sss)/\pi$, are significantly enhanced with multiplicity even in pp
collisions~\cite{ALICE:2016fzo}. The ALICE results on strangeness enhancement
have \emph{irreversibly} falsified the traditional pp paradigm. For models to
describe the data, one therefore has to allow for significant final-state
interactions that break the picture of a pp collision as an almost incoherent
sum of subcollisions. We stress this here (and return to it later) to
emphasize that we now are in a time period where models are struggling to
explain the data and where one can expect fundamental breakthroughs in our
understanding in the coming years.

The rest of this section is organized as follows. First we will examine how
well pQCD describes \pT-spectra of identified particles at LHC energies. After
this we review experimental measurements of different particle species and their
production rates. Finally we will discuss baryon production.

\subsection{Lessons from hard production}

One typically expects that perturbative QCD can describe particle production at
intermediate to high \pT, e.g., $\pT > \gevc{3-4}$. Perturbative QCD (pQCD)
requires two inputs: parton distribution functions (PDFs) and fragmentation
functions (FFs). The PDFs are known to reproduce several observables, e.g.,
jet production and backgrounds in searches for new physics, so when comparing
pQCD calculations with identified spectra one is mostly testing the identified
FFs.  A big surprise from the LHC has been that it has been very difficult for
pQCD to describe spectra of neutral pions and charged pions, kaons and
protons, see for example Refs.~\cite{ALICE:2017ryd,ALICE:2020jsh}. As
suggested in Ref.~\cite{dEnterria:2013sgr}, one of the problems could be that
in some cases there are sizable non-perturbative effects for $\pT < \gevc{10}$,
something we return to in the next section.  Recently it has been possible to
obtain a new set of FFs that provide a good description of both charged and
neutral pion spectra at LHC~\cite{Borsa:2021ran}. \\ Note that in a string picture one can find large non-perturbative corrections to hadron production, even the partonic production is perturbative and so there is a fundamental tension with the idea of universal FFs~\cite{Norrbin:2000zc}.

\subsection{Review of identified particle spectra and production rates}
\label{sec:pid:spectra}

Where the studies of identified particles shine are when the evolution of
different particle spectra are compared. For this reason, we will focus here on
particle ratios as a function of final-state charged-particle multiplicity, \mult.

ALICE is the experiment at the LHC with the best capabilities for studying
identified particle production due to its PID capabilities, low
magnetic field, and low material budget. ALICE has recently published a long review
of the results from LHC, which also includes soft QCD in pp
collisions~\cite{ALICE:2022wpn}. Here, we focus on a few selected results from
this as well as newer results.

For particle ratios as a function of \pT, ALICE has found that at low and
intermediate \pT there can be a substantial evolution with \mult in pp
collisions. This is particularly true for ratios of baryons to mesons,
e.g., $p/\pi$ and \LA/\KZ, while some ratios, such as $K/\pi$, are almost
independent of \pT, see for example Ref.~\cite{ALICE:2018pal}. The
baryon-to-meson ratios have a multiplicity dependence for $\pT <
\gevc{10}$. Some of this multiplicity dependence at low and intermediate \pT is expected from both CR and radial flow,  c.f.\ Eq.~\ref{eq:boost}, and this is likely why it has been difficult to explain identified spectra with pQCD and also why one can find a good set of FFs for pions, due to their low mass. 
However, it is not clear that radial flow or color reconnection can explain all the changes
in particle ratios with multiplicity at intermediate \pT and therefore it could be that one has to include quark recombination effects~\cite{Fries:2003vb}. In
recombination models, baryons (mesons) ``combine'' from three (two) co-flowing
quarks and so baryons naturally end up with significantly higher \pT than
mesons. This could also explain why it is mostly baryon-to-meson ratios
that are modified. For completeness, we note that this mechanism could also play
a central role in understanding the production of heavy flavor baryons, see
for example Refs.~\cite{ALICE:2022exq,LHCb:2023wbo}.  It will therefore be interesting to see if one can validate or falsify this idea in the coming years.

\begin{figure}[h]
\includegraphics[width=0.9\columnwidth]{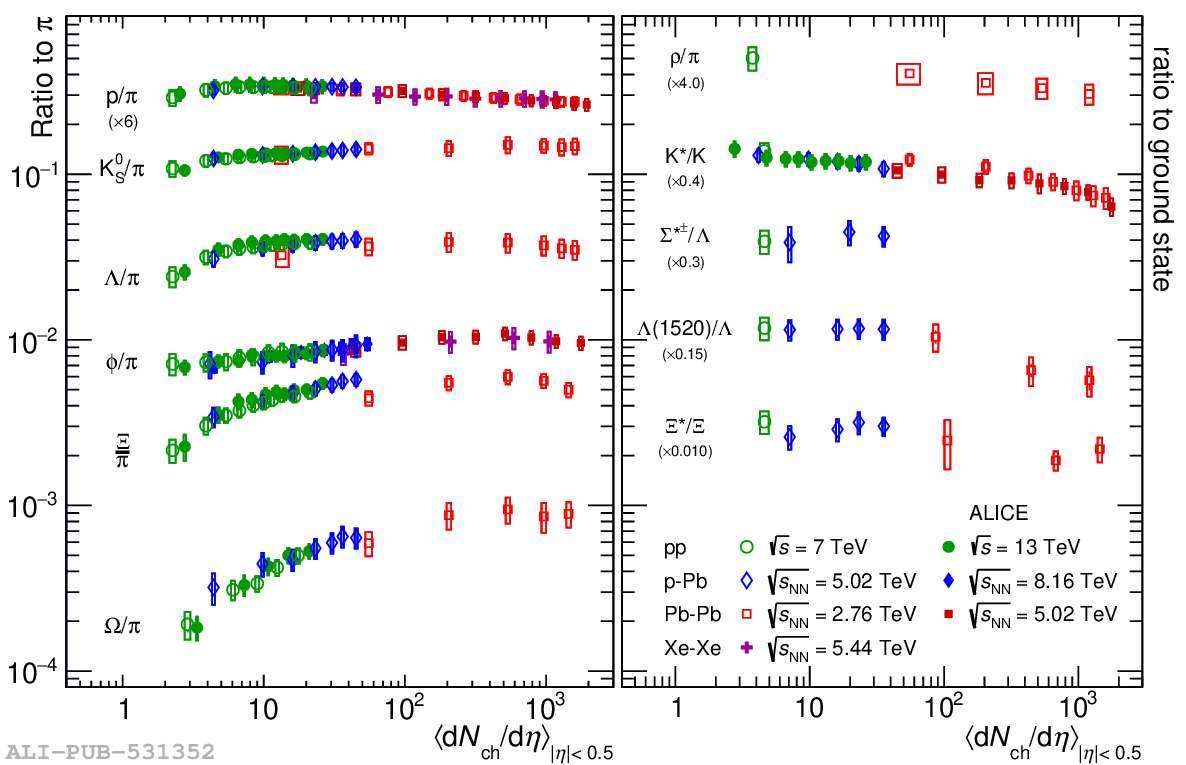}
\caption{The \pT-integrated yield ratios to pions ($\pi^+ + \pi^-$), left, and
  \pT-integrated yield ratios between resonance and corresponding ground
  state, right, are shown as a function of \mult measured in $|\eta| <0.5$
  demonstrating universality across collision systems and beam
  energies. Figure taken from~\cite{ALICE:2022wpn}.}
\label{fig:ratios_vs_mult}
\end{figure}

The ALICE results that have had the biggest impact on soft QCD are the ratios
of integrated particle spectra where it has been found that they only depend on
\mult, irrespective of collision system (pp, pA, and AA) and beam energy, as shown in Fig.~\ref{fig:ratios_vs_mult}. For the cases studied, it has even been found that \pT-differential ratios have this universal scaling for a given
\pT~\cite{ALICE:2018pal}.

It is important to stress here that one has to be careful with how one defines
multiplicity because it is easy to introduce ``selection''  and ``jet'' biases, which
are both caused by \emph{local} fluctuations in the \emph{charged} particle density. These
biases can affect both the measurements of the identified particles and the
\mult. For these studies, the multiplicity selection is done using forward
detectors while the identified particles and the \mult are measured at
mid-rapidity to avoid these biases. Understanding and controlling these and
similar biases is a big challenge in ongoing and future experimental
investigations using new observables.

Returning to Fig.~\ref{fig:ratios_vs_mult}, one notices that the main changes in particle ratios
with \mult are for the strange hadrons including the \PHI meson, and in
particular for the multi-strange baryons. This is commonly referred to as
strangeness enhancement. Several measurements have been carried out to pinpoint the
origin of the strangeness enhancement. By measuring particle ratios inside and
outside of jets it has been established that the strangeness enhancement is
mostly produced in the underlying event~\cite{ALICE:2022ecr}. It is
challenging to constrain precisely the strangeness enhancement for hard jets as it is
expected to be
mainly located at low \pT where the underlying event dominates. There are
indications that at least for ``jets'' with intermediate-\pT leading
particles one can observe an enhancement~\cite{ALICE:2024zxp}. 
A modified version of transverse spherocity, designed to minimize the biases
discussed above, has been used to study particle production in high
multiplicity events~\cite{ALICE:2023bga}. In this study it was found that one can reduce the
strangeness enhancement for fixed \mult by selecting ``jet-like'' events. On the
contrary, it is not really possible to enhance the strangeness production significantly,
which suggests that the bulk of high-multiplicity events are dominated by soft
physics. This can also help explain the
universal scaling observed in Fig.~\ref{fig:ratios_vs_mult} because it suggests that the measurements in all cases are dominated by soft physics and it is only if one selects hard processes or events dominated by hard processes that one can break this scaling.

Let us now focus on the models. The pp models that can describe the
strangeness enhancement have all had to implement some form of final state
interactions. \\
In \pythia, more complicated forms of color reconnection has been
implemented. Where ordinary color reconnection is important for describing the
rise of \meanpT with multiplicity and the radial-flow-like effects in the data, this new color reconnection is centered around the formation of
junctions~\cite{Christiansen:2015yqa}, which are more likely to produce
baryons, and ropes~\cite{Bierlich:2014xba}, which are more likely to produce
strange hadrons due to an increased string tension. The current biggest
challenge for \pythia is that, to explain the strangeness enhancement, one ends
up with a general baryon enhancement due to the junctions that is not
observed for protons, c.f.\ Fig.~\ref{fig:ratios_vs_mult} and
Ref.~\cite{ALICE:2020nkc}.  \\ \herwig has successfully described the
strangeness enhancement by improving the way gluons fragment to
$s\overline{s}$ and allowing three \qqbar clusters close in phase space to reconnect to
form a baryon-antibaryon pair~\cite{Gieseke:2017clv}. While these new mechanisms
allow \herwig to reproduce the strangeness enhancement, they predict enhanced
strangeness production in transverse-spherocity-selected jet-like events,
contrary to what is observed in the data~\cite{ALICE:2023bga}. The reason for
the enhancement in \herwig appears to be the many mesons having parallel
momentum vectors in jet-like events, resulting in enhanced ``baryon
reconnection''. As the ``baryon reconnection'' is reminiscent of
quark recombination discussed above, this suggests that transverse spherocity
analyses can be used to test these ideas if they can be implemented
in generators. \\
When one introduces QGP formation, the change in particle ratios can be explained in two
ways:\\ \epos has a dense QGP-core and a dilute
pp-model-like corona and the change with multiplicity is mainly driven by
the growing importance of the core~\cite{Pierog:2013ria}. Recently, a detailed analysis of the particle ratios has been done in \epos 4~\cite{Werner:2023jps}. There one also tested the effect of hadronic scattering, which is found to have very little effect in pp collisions. The paper interestingly points out that while there is a continuity of particle ratios between pp and AA collisions, the \meanpT in AA collisions is significantly lower than in pp collisions. This is expected in \epos because the core is smaller in pp collisions for the same \mult and therefore expands more leading to larger radial flow.\\
Alternatively, assuming full QGP formation in all pp collisions, one
can describe the change with multiplicity as the transition from a canonical
to a grand-canonical thermal-model description, which can be done in many ways and with and
without full strangeness equilibration, see for example
Refs.~\cite{Vislavicius:2016rwi,Vovchenko:2019kes,Cleymans:2020fsc}. In these
models, strangeness is not enhanced in large systems (AA) but canonically
suppressed in small systems (pp). There are three challenges to these full QGP
models in small systems and the proposed idea of strangeness suppression. 1)
If strangeness is not equilibrated in a small system can we then really talk about a suppression? 2) In the basic
version of the statistical thermal model, the \PHI is expected to be enhanced
in small systems relative to other particles because it has $B=0$, $S=0$, and
$Q=0$ and therefore is never canonically suppressed. However, as can be seen
in Fig.~\ref{fig:ratios_vs_mult} the ratio $\PHI/\pi$ increases with \mult in
pp collisions and so one likely has to treat the \PHI as doubly strange to
explain this~\cite{Vovchenko:2019kes}. 3) A potentially bigger problem is that
none of these models are implemented as full event generators and so one can
really only evaluate integrated ratios and not handle correlations or
biases, which are important in pp collisions. It would be great if more
resources in the field would go into formulating a full generator built around
this idea. Without event generators, it is almost impossible to see how one
can validate or falsify the interesting and perspective-changing idea of strangeness
suppression in small systems.

\subsection{Baryon production}

Recently a new type of measurement has become available at the LHC where one
directly studies the microscopic balance of quantum numbers, $B$, $S$,and $Q$,
for \XI baryons by other hadrons~\cite{ALICE:2023asw}. The measured microscopic
balance has turned out to challenge all models as it does not depend
significantly on multiplicity indicating that the microscopic production of
\XI is more or less the same \emph{independent} of the yield enhancement. For
example, in \pythia the strangeness enhancement is mainly explained by junction formation, which can be observed in the balance results as one is much more likely to balance the baryon
number of a \XIM by a $\overline{p}$ than for the default diquark model, which
requires one $\overline{s}$ in the balancing antibaryon. However, junction
formation naturally has a multiplicity dependence as it depends on
the string density so that one
expects changes in the \XI
balancing with multiplicity, which is not observed in the data. \\
As this type of measurement allows one to constrain the microscopic production
mechanisms there is hope that similar measurements of other particle species can lead to
new breakthroughs in our
understanding of baryon production and hadronization in general.\\

An effect that is observed in data, but does not appear to be in any of the
models, is that $B-B$ (and $\overline{B}-\overline{B}$) correlations are
suppressed close in phase space~\cite{ALICE:2016jjg}. As the strangeness enhancement is mainly observed for strange
baryons one could worry that we are overlooking essential pieces of
information that could improve our understanding of baryon production. It would therefore also be interesting if CMS or ATLAS, which can measure \LA-\LA correlations over a much broader rapidity range than ALICE, would also study these correlations. 

\section{Collectivity: Long-Range Correlations and Elliptic Flow}
\label{sec:collectivity}

In large collision systems, the observation of anisotropic collective flow has been one
of the most transformative measurements in the field of heavy-ion physics, see
Ref.~\cite{Heinz:2013th} for an overview. The observation of collectivity in pA and pp collisions, see for example
Refs.~\cite{Nagle:2018nvi,Grosse-Oetringhaus:2024bwr} for overviews, both
raises questions about the origin of the flow and offers new possibilities in
these fields. As the reference system in this case is AA collisions, we will
first start by describing the physics there, see Ref.~\cite{Heinz:2013th} for
details.

The idea of flow is easiest to understand in the hydrodynamic paradigm. In hydrodynamics, the expansion of the QGP formed in AA collisions is driven by pressure gradients. As semi-central collisions have an anisotropic “elliptic” spatial distribution, the hydrodynamic expansion will lead to a momentum anisotropy that has its major (minor) axis along the minor (major) spatial axis. In general one can describe the final state by a Fourier series:
\begin{equation}
  \label{eq:flow}
  \frac{dN}{d\varphi} \approx \frac{N}{2\pi} \left ( 1 + 2 \sum_{n=1}^{\infty} v_n
    \cos (n (\varphi - \Psi_n)) \right )
\end{equation}
where $\varphi$ is the azimuthal angle, $\Psi_n$ is the $n$-th order symmetry
plane and $v_n$ are the flow coefficients. The elliptic flow, $v_2$, dominates
AA collisions and is mainly driven by the average geometrical overlap and
$\Psi_2$ is approximately the azimuthal angle of the impact-parameter vector. The
triangular flow, $v_3$, has turned out to be driven by event-by-event
fluctuations, which means that $\Psi_3$ is independent of $\Psi_2$.

Flow measurements have transformed the field of heavy-ion collisions in the following ways:
\begin{itemize}
\item The large magnitude of the elliptic flow requires a perfect liquid with
  a shear-viscosity-to-entropy density close to the universal value for a
  large class of strongly interacting quantum field
  theories~\cite{Kovtun:2004de}. This has falsified the idea of a weakly coupled QGP and established the strongly interacting QGP, sQGP.
\item The observation of large triangular flow indicates that event-by-event
  fluctuations in the initial state are important and can be probed with
  anisotropic flow and event-by-event flow fluctuations. This has even
  provided some sensitivity to the magnitude of subnucleonic
  fluctuations~\cite{Schenke:2012wb} that could be relevant for our
  understanding of pp collisions.
\end {itemize}
The important thing to understand is that a perfect liquid has as little
entropy generation as possible, i.e., almost no diffusion or dissipation, and
so it preserves information from the stages of the collisions that we
cannot otherwise easily probe. This has allowed Bayesian analyses to constrain the equation of state and many other properties of the
QGP~\cite{Bernhard:2016tnd}.

\begin{figure}[h]
\begin{minipage}{0.46\textwidth}\centering
\includegraphics[width=0.9\columnwidth]{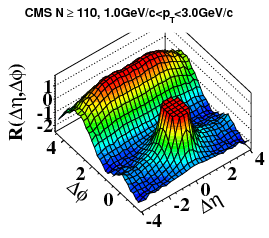}
\end{minipage}
\begin{minipage}{0.54\textwidth}\centering
\includegraphics[width=0.9\columnwidth]{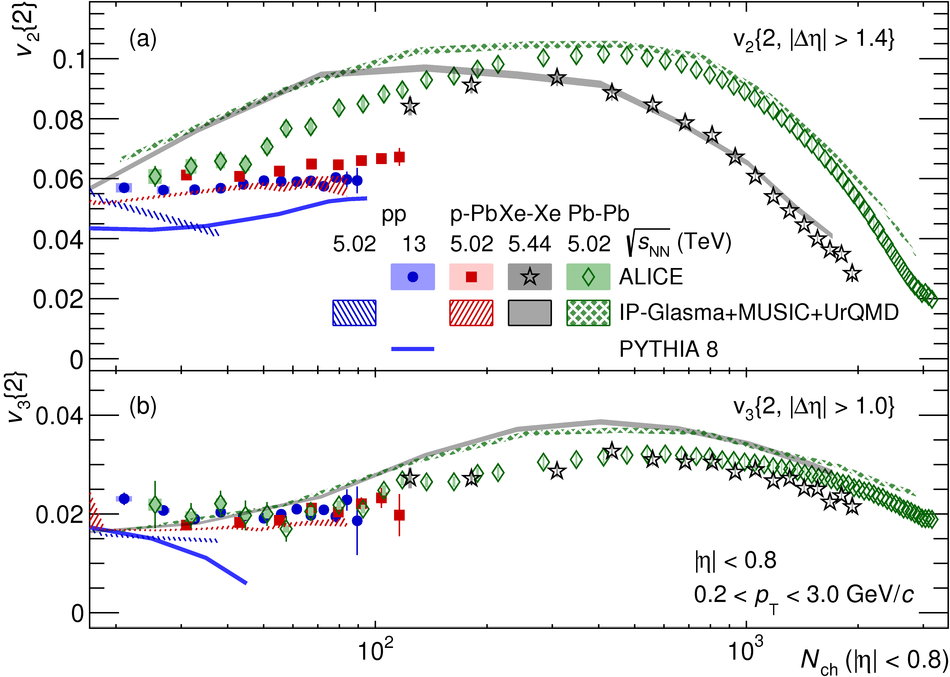}
\end{minipage}
\caption{Left: the first observation by CMS of the near-side ridge in high-multiplicity pp collisions. Figure adapted from~\cite{CMS:2010ifv}. Right: measurements of $v_2$ and $v_3$ by ALICE for pp, pA, and AA collisions using measured using two-particle cumulants with a pseudorapidity separation $|\Delta \eta| > 1.4$ and 1.0, respectively, chosen to suppress nonflow contributions. Figure adapted from~\cite{ALICE:2019zfl}.}
\label{fig:vtwo}
\end{figure}

In small systems, flow is difficult to observe, which can be understood in the
following way. The correlations imprinted by flow are weak but present between
all particles (global), while non-flow correlations, such as those resulting
from (mini)jets and resonance decays, are strong but only involves few
particles (local). For this reason, flow-induced correlations dominate for
high multiplicities (AA collisions) while non-flow correlations dominate for
low multiplicities (pp collisions), which also makes it impossible to measure
the symmetry planes $\Psi_n$ in pp collisions. In a recent
review~\cite{Grosse-Oetringhaus:2024bwr}, the many different ways to extract
flow in small systems have been described in detail. Here we shall just go
through some of the main observations of flow in small systems at the LHC. \\
The first observation was the near-side ridge in very high-multiplicity pp
collisions observed by CMS in 2010~\cite{CMS:2010ifv}, as shown in \textbf{Figure~\ref{fig:vtwo} left}. This was followed by
the observation of the double ridge in p-Pb collisions by
ALICE~\cite{ALICE:2012eyl} and ATLAS~\cite{ATLAS:2012cix}. Since these first results, a lot of work has gone into refining the techniques to measure flow and ALICE has recently conducted a
systematic study of flow across all systems~\cite{ALICE:2019zfl}. The study found
that the flow coefficients $v_2$ and $v_3$ are similar for different
systems, pp, pA, and AA, when compared at the same \mult in the
low-multiplicity region, see \textbf{Figure~\ref{fig:vtwo}}. In this region, one expects  the initial state geometry to be driven by fluctuations and therefore to be similar for all systems. The observation that the flow coefficients are the same therefore suggests that the origin of the flow is the same. It is worth to point out that the results with $v_2$ were validated in the same paper using using flow cumulants with two and
three subevents~\cite{Jia:2017hbm}, which is one of
the most precise measurements of flow.

Now let us turn to discuss the origin of flow in small systems. For some time
it appeared that the flow in small systems might be produced by initial-state
effects, as predicted by the Color Glass Condensate
(CGC)~\cite{Dusling:2017dqg}, but this has since then been disfavored both
experimentally and theoretically.  Experimentally, RHIC carried out a
spectacular program of p-Au, d-Au, and $^3$He-Au collisions with which they
could demonstrate that initial-state effects could not explain the
flow~\cite{PHENIX:2018lia}. Theoretically, it was recently realized that when
one combines CGC calculations with final-state effects, needed to produce a
hydrodynamic medium in larger systems, the long-range initial-state flow
correlations disappear~\cite{Schenke:2022mjv}. This means that the only
hypothesis that is currently valid is that flow in small system is to a large
degree due to final-state interactions.  As this is also what we expect to be
the case for the strangeness enhancement discussed in the previous section,
the two observations are consistent.

The main open questions when it comes to flow are the following ones:
\begin{itemize}
\item What are the smallest systems that can flow?
\item How can we theoretically understand how flow arises in very small systems?
\end{itemize}

The surprising observation in pp collisions has been that there does not
appear to be an onset of flow with
multiplicity~\cite{Grosse-Oetringhaus:2024bwr,ALICE:2019zfl}. In fact,
flow-like signatures have been observed even for multiplicities that are smaller
than those of MB pp collisions~\cite{ALICE:2023ulm}. This has shifted the
question of the possible onset of flow to different collision systems. Using
LEP data at the $Z$ peak, it has been found that flow does not appear to be
present in $e^+e^- \rightarrow q\overline{q}$
collisions~\cite{Badea:2019vey}. 
However, results for higher LEP beam energies, above the $e^+e^- \rightarrow W^+W^-$ threshold ,
suggest that flow-like correlations are present there~\cite{Chen:2023njr}. Other types of
systems that have been studied are ultra-peripheral collisions (UPCs) at LHC and
$e$p collisions at DESY. In these systems, one expects that the $\gamma^*$ will
interact hadronically for low $Q^2$ and so they are relevant to compare with.
ATLAS has observed flow in UPC $\gamma^*$-Pb events~\cite{ATLAS:2021jhn} and
CMS has reported non-zero $v_2$ in $\gamma^*p$ events~\cite{CMS:2022doq}, but
finds that the magnitude is consistent with model predictions that have no
collective effects. ZEUS and H1 have reanalyzed old data both for low and high
$Q^2$ but neither ZEUS~\cite{ZEUS:2019jya,ZEUS:2021qzg} nor H1~\cite{H1}
observes any signatures of collective flow. To use these systems to establish the onset of flow it would be good to see
investigations of both UPC systems by a single collaboration and even
discussions between analyzers at LHC and HERA to ensure that the different results are not caused by differences in analysis choices.

One of the biggest questions for small system has been the following: how can
we form an equilibrated medium in these small systems? This is a very active
and difficult theoretical topic with many breakthroughs. The first major
breakthrough, resulting from efforts on several fronts, was the realization
that hydrodynamics does not require thermalization but only hydrodynamization,
see Ref.~\cite{Strickland:2024moq} for a recent review. In weakly-coupled
kinetic theory one can study out-of-equilibrium systems in detail and build up
a better understanding of hydrodynamization. There, one finds that systems
first reach kinetic equilibration, then chemical and finally thermal
equilibrium, see for example Ref.~\cite{Kurkela:2018xxd}, and that the system
can already be described by hydrodynamics when it is in kinetic
equilibrium. When one estimates the system size (multiplicity) at which one
can have hydrodynamic flow, it is found to be significantly bigger than the
range of multiplicities where flow-like effects are observed in pp
collisions~\cite{Kurkela:2018xxd,Ambrus:2022qya}. It turns out that one can
generate flow-like signals that are not hydrodynamic at lower multiplicities. An
example of this is the escape mechanism where the flow-like signal is induced
by an azimuthally asymmetric escape probability and only requires one
interaction on the average~\cite{Kurkela:2018ygx}. In fact, one can generate
flow-like signals in this way even in realistic QCD kinetic
theory~\cite{Kurkela:2021ctp}. This is a fantastic result, but unfortunately
it is sometimes interpreted as an indication that everything flows and that
perfect-liquid-like behavior in pp collisions is not very special. Instead,
the community needs to develop generators that can be compared
apples-to-apples with data to see if this proposed mechanism is indeed a
viable explanation. We think this is urgently needed as there are reasons to
be skeptic: no onset of flow is observed in pp collisions while any
weak-coupling calculation naturally has a finite mean free path. This makes it
likely that strong-coupling physics is needed to explain the observed flow in
pp collisions, which is difficult to calculate, but could make the flow in pp
collisions perfect (vanishing mean free path).  We therefore encourage heroic
efforts to continue progress on this side as well.

\subsection{New directions}

CMS has recently reported flow-like effects inside very high-multiplicity jets
in pp collisions~\cite{CMS:2023iam}. This is unexpected and it would therefore
be good if other experiments could confirm the results. It would also be
important to understand the partonic origin of these very high-multiplicity
jets and to understand the new coordinate transformation involved in the
measurement better.

There are two other directions that we think should be further explored in the case of flow in
small systems. In AA collisions, one observes significant jet quenching
meaning that jets lose energy as they propagate through the sQGP. As this is
a final-state interaction and since final-state interactions are present in pp
collisions, one would expect that jet quenching could also occur
there. However, so far there are no indications of jet quenching in pp and pA
collisions, see for example Refs.~\cite{ALICE:2023plt,ATLAS:2022iyq}. This
could be due to the magnitude of jet quenching being too small or the statistical and systematic limits of available measurements too large, but
it could also mean that there is no jet quenching in pp collisions. If this is
the case then this must be an important part of understanding the origin of
flow and strangeness enhancement in these systems and we therefore encourage
theorists to explore this possibility.

The second direction is the possibility to use flow in a similar way as in AA
collisions to learn about the microscopic event-by-event fluctuations in pp
collisions, which could provide an alternative way to constrain the UE. There
are for example indications that the flow is very sensitive to how the proton ``geometry''
is implemented in pA collisions~\cite{Mantysaari:2017cni}. In a recent paper
it was even suggested that the sign of the elliptic flow in pp collisions can
be used to image the relevant degrees of freedom in the final
state~\cite{Bierlich:2024lmb}, i.e., that if $v_2$ is in fact negative then
this would be an indication that the flow is not hydrodynamic but could be the
result of Lund strings ``shoving'' each other or long-range partonic
interactions. This raises a broader question related to flow in small systems:
can we ignore the size of the fundamental degrees of freedom in these systems
and neglect the precise interaction length as is done in hydrodynamic models?
We hope for further phenomenological developments along these lines to provide
new directions and to guide experimental investigations.

\section{Conclusions and Outlook}
\label{sec:outlook}

At first sight, it may seem that a fundamental tension exists in soft QCD physics. On one hand, there is no direct evidence of saturation effects, and proton-proton collisions at 13 TeV are widely considered far from the Froissart-Martin bound. This suggests that partonic densities remain below the threshold for significant recombination effects. On the other hand, hydrodynamic models, which describe the evolution of dense and coherent systems, are frequently invoked to explain phenomena such as collective flow observed in multi-particle correlation measurements. This duality—where collisions are seen as both dilute enough to avoid saturation and dense enough to exhibit collective behavior—poses a significant challenge. Resolving this contradiction requires a deeper understanding of the interplay between partonic dynamics, saturation, and coherence in small systems.

The traditional pp paradigm that dominated soft QCD physics prior to the LHC
was one that in some sense is derived from perturbative QCD. In particular for
\pythia, the soft and hard physics was modeled similarly based on the idea of Jet Universality. With LHC,
soft QCD physics has irreversibly changed as models have to add final-state
interactions, such as color reconnection, between subcollisions and hadronization. One could add
here that to some degree,
even perturbative QCD has changed with the observation of charm and bottom
baryons being enhanced over mesons with respect to $\ee \rightarrow
\qqbar$. Importantly, these effects were discovered at LHC but we refind them in the data at lower beam energies. The three
main reasons for why these effects were discovered at LHC were the extended multiplicity reach of the LHC
beam energies (as the effects grow with multiplicity), the larger focus on
particle identification by in particular ALICE, and the new paradigm that AA
collisions can be a reference for the search for QGP-like effects in pp collisions.

The breaking of Jet Universality raises new questions that could be ignored
in the traditional, dilute pp paradigm. What are the relevant degrees of freedom after
the subcollisions? How do they microscopically form/equilibrate and interact?  It therefore appears that some aspects of the heavy ion paradigm could be applicable to pp collisions. Unfortunately, it seems that while the experimental questions of soft QCD in pp
and AA collisions have merged, the theoretical/phenomenological approaches are
living in parallel worlds. This is very unfortunate because the pp
generator community have a huge expertise on the event-by-event fluctuations
that are essential to model pp collisions, while the AA community have an
expertise on these out-of-equilibrium processes
that needs to be included when you have final-state interactions, see for
example Ref.~\cite{Berges:2020fwq} for an overview. A stronger knowledge
transfer and collaboration between these two groups could benefit the
understanding of soft QCD at LHC tremendously. Furthermore, having
AA-inspired phenomenology available in an open MC generator form that can be
directly compared with the data (including biases due to, for example,
multiplicity estimation) or used to devise tests to validate or falsify the
proposed mechanism is urgently needed. To facilitate progress on this side, one should
investigate if some of these mechanisms can be implemented as add-ons to
existing pp generators.

On the experimental side there is a need to make more results available in a
format that is easy to compare with pp generators, e.g., in \rivet. This would
facilitate the tuning of models and also in general ensure the best
apples-to-apples comparison with data. Tools have been developed to facilitate
the new QGP-like analyses in \rivet~\cite{Bierlich:2020wms}, but there are
still not many examples and it is therefore not always easy to implement the
new analyses. A dedicated concentrated expert effort is needed to get this going.

Another important point to raise here is that the potential improvements of
these efforts on the LHC discovery program needs to be investigated. Since
most of the discovery program is focused on high \pT and most of the soft
physics at low \pT it is not so obvious what can be gained. At the same time, this in some sense makes it even more important to carefully examine if this is indeed the case to not miss out on a big opportunity by the tremendous progress on our understanding of soft QCD at LHC.\\

Although dilute proton-proton (pp) collisions and dense heavy ion collisions may seem quite distinct, and one must be cautious when applying concepts from one to the other, we believe that addressing unresolved questions in soft QCD physics can benefit greatly from the exchange of ideas and analytical techniques between these areas. Such cross-fertilization is essential for improving the accuracy of models used to describe physics in unexplored energy ranges, such as those anticipated at future accelerator facilities.

Finally, we hope we have highlighted in this review that there is no reason to expect that the age of
experimental exploration and digging for gold at the LHC is over.
\begin{quote}
\textit{All that is gold does not glitter,\\
Not all those who wander are lost}\\
\hspace*{2.7cm} J.R.R.\,Tolkien
\end{quote}

\section*{DISCLOSURE STATEMENT}
If the authors have noting to disclose, the following statement will be used: The authors are not aware of any affiliations, memberships, funding, or financial holdings that
might be perceived as affecting the objectivity of this review. 

\section*{ACKNOWLEDGMENTS}
PC would like to thank the participants in the ``QCD challenges from pp to AA
collisions'' International Workshop in Münster in September 2024 for inspiring
presentations and discussions.\\

Support from the following research grants are gratefully
acknowledged. Vetenskapsrådet contract 2021-05179 (PC). Knut and Alice
Wallenberg foundation contract number 2017.0036 (PC).

%

\bibliographystyle{ar-style5}
\bibliography{review_bibliography}

\end{document}